\documentclass[11pt,preprint2]{aastex}
\usepackage{natbib}
\usepackage{graphicx}
\usepackage{url}

\begin{document}
\title{Search for TeV Emission from Geminga by VERITAS}
\author{Gary Finnegan for the VERITAS Collaboration
}
\affil{Department of Physics and Astronomy\\ University of Utah, Salt Lake City, UTAH 84108-0830}
\author{VERITAS Collaboration}
\affil{for a full authors list see\\ http://veritas.sao.arizona.edu/conferences/authors?icrc2009}

\begin{abstract}
The Geminga gamma ray source was first detected by the SAS-2 gamma-ray satellite observatory and the COS-B x-ray satellite observatory, and has been identified as a radio-quiet pulsar associated with a 300,000 year old supernova remnant. Geminga is one of the brightest GeV sources as seen by the Large Area Telescope on board the Fermi gamma-ray satellite observatory. A ground based detection was accomplished by the Milagro extensive air shower array at energies greater than 20 TeV. During 2007 VERITAS performed observations to search for TeV gamma ray emission from the Geminga pulsar and the region near Geminga. In this paper, we describe these measurements and the new analysis of these observations. \\
\end{abstract}

\section{Introduction}
\begin{figure}[tttt]
   \includegraphics[height=.3\textheight,angle=270]{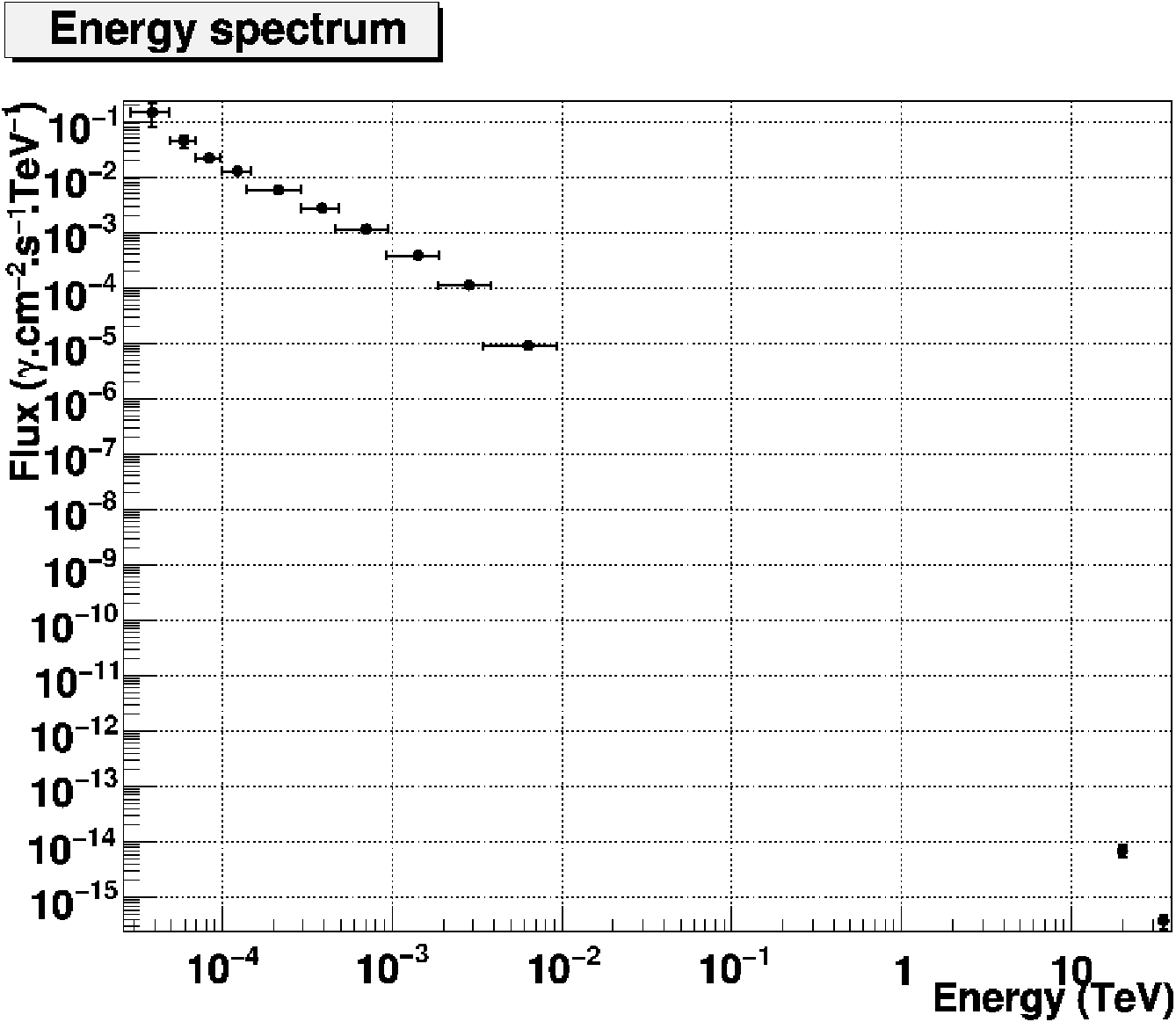}
  \caption{The differential energy flux of the Geminga pulsar from EGRET \citep{Fierro1998} (top
    left hand points) and the differential energy flux of Geminga from Milagro
    (lower right hand corner) \citep{Abdo2007} \citep{Abdo2009} shows the energy
    gap between the two measurements.}
  \label{spectrum}

\end{figure}

Geminga is a relatively close pulsar approximately 200 parsecs away \citep{Caraveo1996} \citep{Faherty2007}.  It was first detected in high-energy ($> 20 MeV$) gamma rays by the SAS-2 satellite \citep{Fichtel1975}. In 1977 the COS-B x-ray satellite was launched. \citet{Hermsen1977} reported a detection of x-ray radiation from the COS-B data of the same region as \citet{Fichtel1975}. \citet{Bignami1983} were able to determine that Geminga was a neutron star from the COS-B data. No pulsations from the star were detected in the SAS-2 or COS-B data. A 237 millisecond period of pulsed gamma rays was first detected by the EGRET telescope aboard the Compton Gamma Ray Observatory satellite \citep{Halpern1992}. Shortly after the report of Halpern and Holt the Egret gamma-ray observatory measured a period derivative of $1.9521712 \times 10^{-13}$ [Hz $s^{-1}$] \citep{Bertsch1992}, allowing an extrapolation of Geminga's ephemeris back to the era of the SAS-2 observations. Subsequently 237 millisecond pulses were now detected in the archival SAS-2 and COS-B data.  From the pulsed analysis of Geminga a characteristic age was calculated to be $\sim 300,000$ years old. The only pulsed radio emission detection of Geminga has come from the Pushchino Radio Astronomy Observatory at 102.5 MHz from three groups, \citet{Malofeev1997}, \citet{Kuzmin1997}, and \citet{Shitov1998}. No other radio frequencies have been detected from Geminga. More recently the Large Area Telescope on the Fermi gamma-ray space satellite observatory (Fermi LAT) has also detected  high energy gamma rays from Geminga \citep{Abdo2010}. Very high energy gamma rays (TeV) have been detected by the Milagro extensive air shower array that are positionally coincident with Geminga at energies greater than 20 TeV \citep{Abdo2009}. This leaves a gap in the energy spectrum of Geminga from the GeV range to 20 TeV, see figure \ref{spectrum}. Imaging atmospheric Cherenkov telescopes (IACTs) such as VERITAS have a sensitivity from 100 GeV to greater than 30 TeV that will cover the energy gap between space-based detectors and the Milagro ground-based extensive air shower array. 

\section{VERITAS}
The VERITAS IACT array is located south of Tucson, Arizona, at the Fred Lawrence Whipple Observatory. Each telescope is equipped with a 499 photomultiplier tube camera with a 500 mega-samples per second flash ADC readout system.  The optical reflector of each telescope is twelve meters in diameter and uses a Davies-Cotton design. VERITAS IACTs have an energy range of 100 GeV to greater than 30 TeV with and energy resolution of 15\% at 1 TeV. The peak effective photon collection area is approximately 100,000 square meters. The angular resolution is 0.1 degrees at 1 TeV with a location accuracy less than 50 arc seconds. VERITAS is capable of detecting the Crab nebula (the standard candle for TeV astronomy) in two minutes and a source with a flux of 1\% of the Crab in less than 50 hours.  VERITAS operates September through July, with an average yield of 750 hours per year of time when the Moon is set, and 100 hours per year when the Moon is above the horizon. 

VERITAS detects the Cherenkov light emitted by an extensive air showers (EASs) that is created when a gamma ray or cosmic ray enters and interacts with the atmosphere. The image parameters of the EAS are then compared with image parameters that are derived from Monte Carlo simulations and comparisons are made to separate gamma-ray showers from cosmic-ray/hadronic showers. Geometric parameters are applied to each event to calculate the arrival direction and the impact distance on the ground of the EAS. For each EAS the number of photo electrons from the photomultiplier tubes is counted. By comparing the shape and impact parameters and the photo electrons produced by the EAS to simulations, the energy of the gamma ray is estimated.

\section{Observations, Analysis and Results}

During November and December of 2007, VERITAS collected 15 hours of data. After quality weather selection 10.4 hours of observations were selected to analyze. All 
observations were taken in wobble mode, where the source is offset 0.5$^\circ$ from the camera center. The average observation was performed at an elevation of 71.4$^{\circ}$. 

The data from the wobble mode observations were analyzed using a point source analysis using the reflected region background method \citep{Aharonian2001} and the ring background method \citep{Berge2007}. A significance is determined using the likelihood ratio method from \citet{LiandMa}.  This type of analysis was done on the Crab nebula which resulted in a significant detection of $\sim 8$ gamma rays per minute and a differential flux at 1 TeV of $(3.63 \pm 0.15)\times 10^{-12}$ {\it photons} $TeV^{-1} cm^{-2}s^{-1}$.

\begin{figure}[tttt]
  \includegraphics[height=.3\textheight,angle=270]{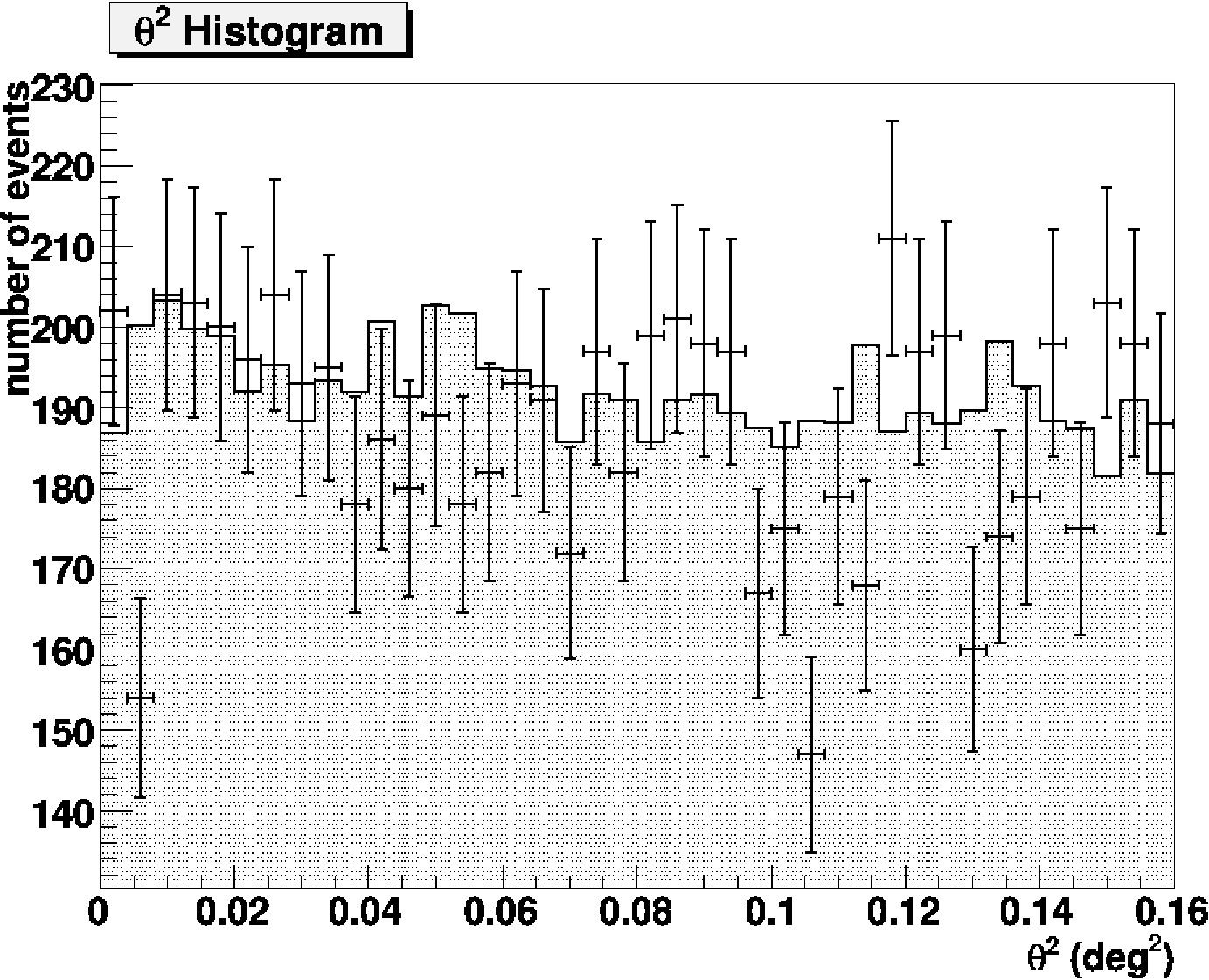}
  \caption{Comparison of ON (points) and OFF (shaded area) events as function of the distance
    from the source center measured in degrees squared
    ($\theta^2$). 
     }
  \label{p2}

\end{figure}

\begin{figure}[tttt]
  \includegraphics[height=.3\textheight,angle=270]{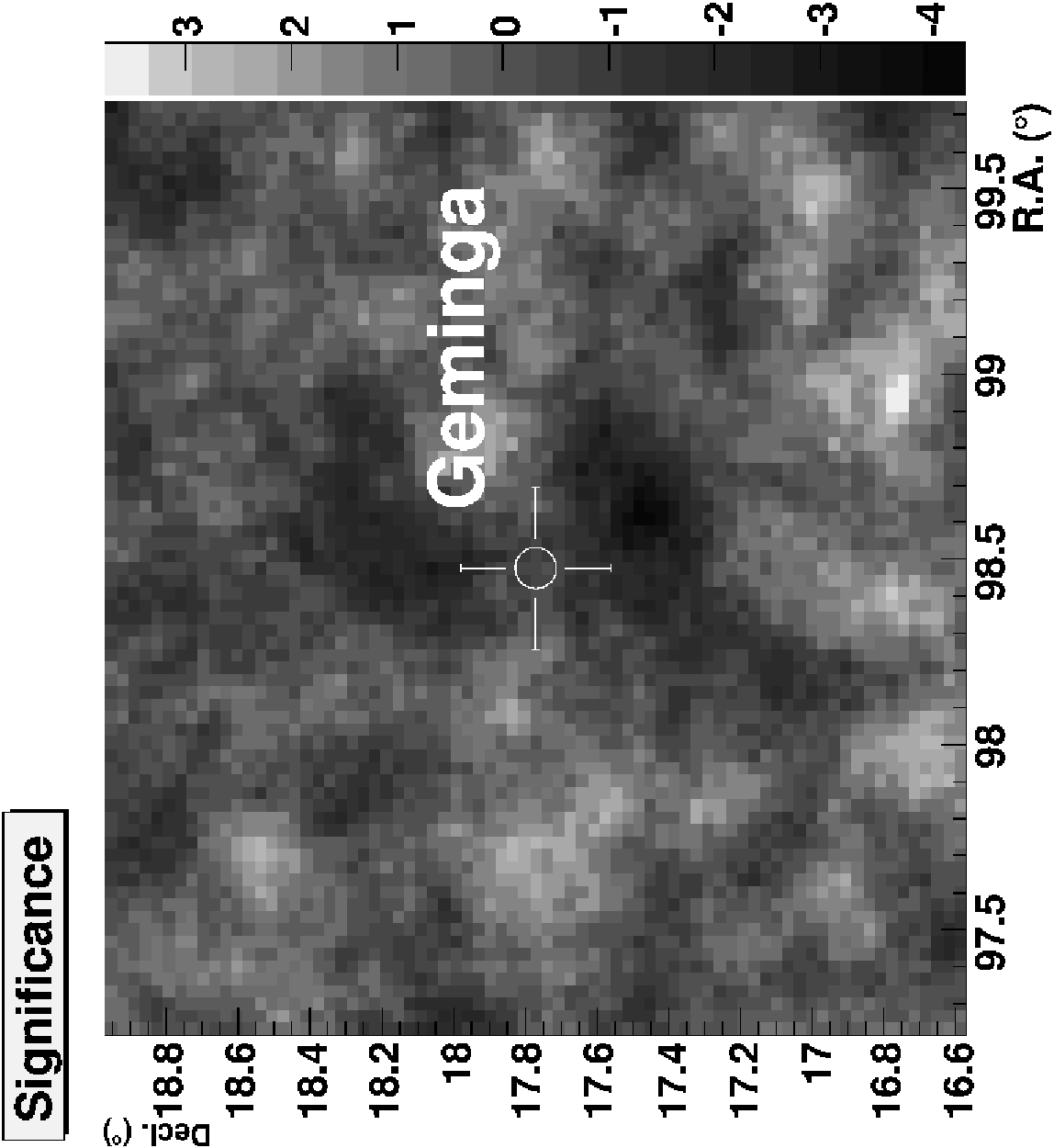}
  \caption{2D significance plot of the Geminga Region }
  \label{p1}
\end{figure}

\begin{figure}[tttt]
  \includegraphics[height=.3\textheight,angle=270]{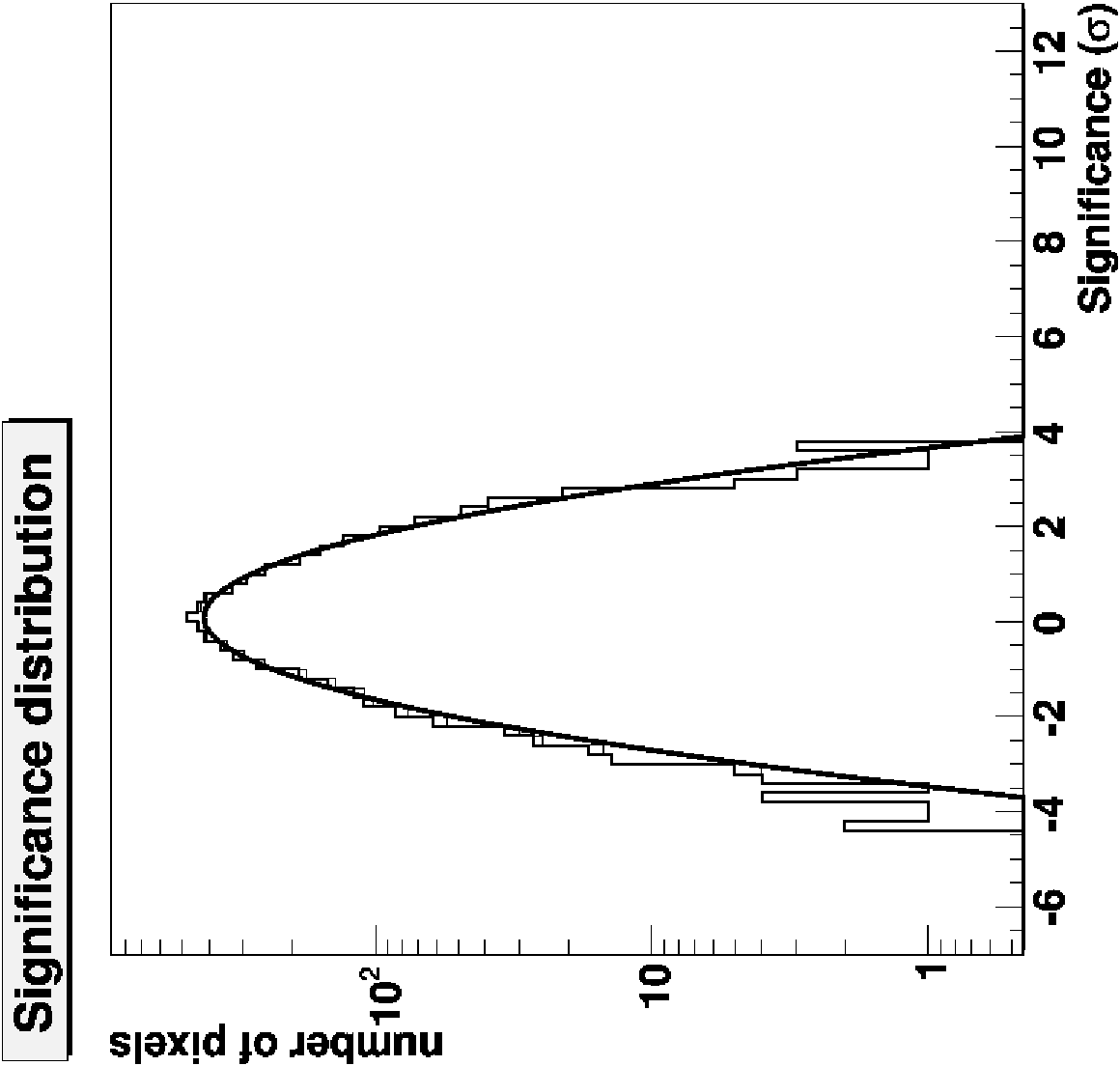}
  \caption{Significance distributions histogram of the 2D significance plot (Figure \ref{p1})}
  \label{p3}
\end{figure}

A steady/point source analysis was first done on the Geminga data. Figure \ref{p2} shows a histogram of the squared angular distribution of the ON (signal) and OFF (background) events recorded around the Geminga source location. The  ON source data points of figure \ref{p2} are consistent with the OFF region (shaded area), and no significant excess was found. Figure \ref{p1} shows a 2-D map of the significance for each map bin surrounding Geminga and figure \ref{p3} is a 1-D histogram of the significance of each bin in the 2-D map. The 1-D histogram of significance is well fit by a Gaussian distribution corresponding to the expected background signal. The 99\% confidence level \citep{Helene1990} limit of the steady/point source analysis for energies above 300 GeV is $< 2\times 10^{-12}\,photons \, cm^{-2}s^{-1}$. 

\begin{figure}[tttt]
  \includegraphics[height=.25\textheight]{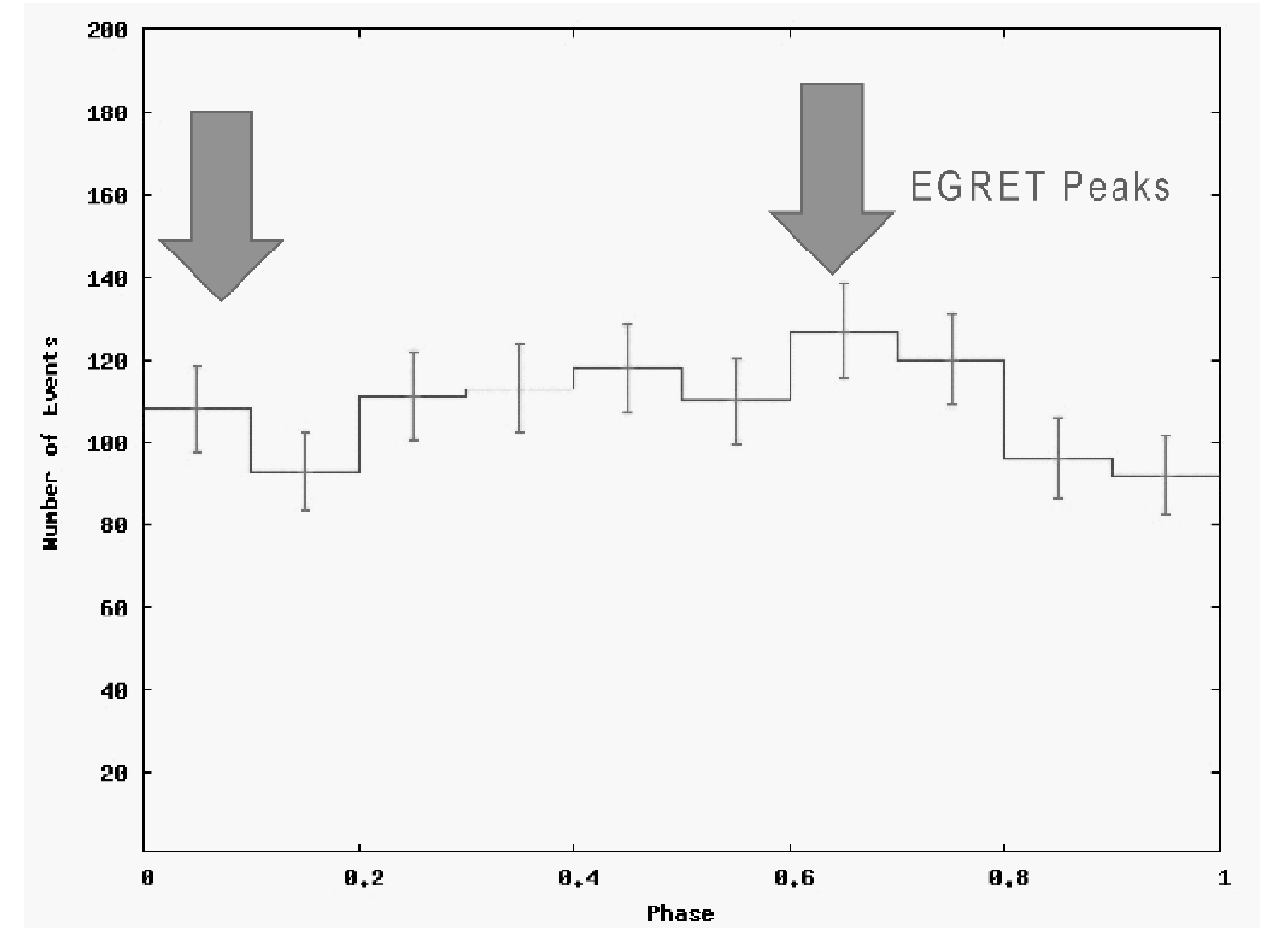}
  \caption{Histogram of the phase distributions of all the ON events. Arrows represent the peak locations in the
    EGRET phase plot \citep{Jackson2005}.}
  \label{p4}
\end{figure}

\begin{figure}[tttt]
\includegraphics[height=.25\textheight]{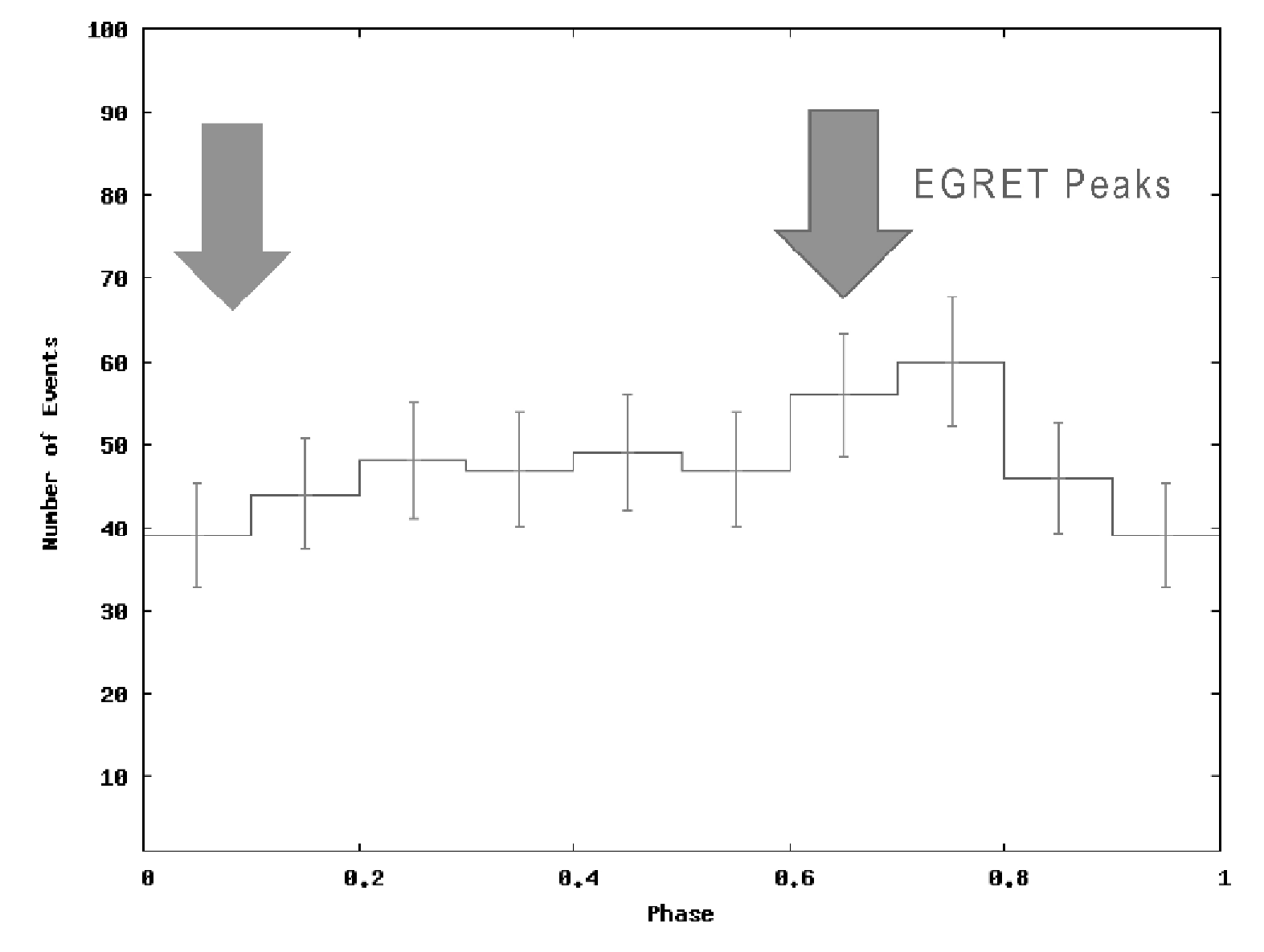}

  \caption{Histogram of the phase distributions of the ON events with an energy
    $< 200$ GeV. Arrows represent the peak locations in the
    EGRET phase plot \citep{Jackson2005}.}
  \label{p5}
\end{figure}

A periodicity analysis which is similar to the steady/point source analysis was done on the same data. The arrival times of each event, after barycentric corrections were applied, were compared to the ephemeris of Geminga based on XMM-Newton and ASCA x-ray space satellite observatories and EGRET gamma-ray observatory observations \citep{Jackson2005}. Figure \ref{p4} shows a histogram of the phase distribution for all ON events in the data. Figure \ref{p5} is similar to Figure \ref{p4} but for energies less than 200 GeV. The arrows represent the peak locations in the EGRET phase plot. No correlation was found with the VERITAS data and the pulsar phase distribution of Geminga. The 99\% confidence level limit of the periodicity analysis is $0.8-1.0 \times 10^{-12}\, photons \, cm^{-2}s^{-1}$.   An extended-source analysis is progress.

\section{Discussion}

Despite higher resolution measurements, the energetic high energy emission from Geminga continues to puzzle astronomers. The proximity of the Geminga pulsar and supernova remnant could explain the possible electron-positron excess as seen by ATIC experiment \citep{Chang2008}. It has also been suggested by \citet{Salvati2008} that Geminga could be responsible for the Milagro cosmic ray anticenter hot spots \citep{Abdo2008}. 

\citet{Caraveo2003} estimated that Geminga is accelerating electrons to energies greater than 100 TeV with a bow-shock like structure. The accelerated electrons from the bow shock should create high energy gamma rays from either collisions from other particles or inverse-Compton scattering. Milagro has detected an excess of TeV gamma rays of high statistical significance with an extension of $2.6^{+0.7\circ}_{-0.9}$ that is positionally coincident with Geminga \citep{Abdo2009}. This makes the diffuse size of Geminga's extension approximately 5 to 10 parsecs in diameter.

A significant detection of gamma rays is needed from VERITAS in order to differentiate the particle acceleration processes in the Geminga region. If a extended detection is made by VERITAS, a spectral energy distribution of Geminga by VERITAS will bridge the gap of scientific knowledge of Geminga between the Fermi LAT and Milagro detections.

\section{Acknowledgments}
This research is supported by grants from the US Department of Energy, the US National Science Foundation, and the Smithsonian Institution, by NSERC in Canada, by Science Foundation Ireland, and by STFC in the UK. We acknowledge the excellent work of the technical support staff at the FLWO and the collaborating institutions in the construction and operation of the instrument.

\bibliographystyle{apj}                    
\bibliography{apj-jour,ref}

\end{document}